\documentclass[aps,prc,twocolumn, superscriptaddress, showkeys]{revtex4-1}

\usepackage{multirow}
\usepackage{graphicx}  
\usepackage{dcolumn}   
\usepackage{bm}        
\usepackage{amssymb}   
\usepackage{graphicx, epsfig, subfigure}
\usepackage{hyperref}
\usepackage[mathlines]{lineno}
\usepackage{color}
\usepackage{float}
\usepackage{enumerate}
\usepackage{enumitem}
\usepackage{tabularx}
\usepackage{multirow}
\usepackage{amsmath}
\usepackage{tensor}
\usepackage[export]{adjustbox}
\usepackage{upgreek} 

\usepackage{numcompress}
\newcommand{\sqrtsnn}{\ensuremath{\sqrt{s_{\mathrm {NN}}}}}

\newcommand{\npart}{\ensuremath{\langle N_\mathrm{part} \rangle}}

\newcommand{\pT}{\ensuremath{p_\mathrm{T}}}

\begin{document}

    \title{Correlations of Baryon and Charge Stopping in Heavy Ion Collisions} 
    \author{Wendi Lv}
    \affiliation{State Key Laboratory of Particle Detection and Electronics, University of Science and Technology of China, Hefei, Anhui 230026, China} 
    \author{Yang Li}
    \affiliation{State Key Laboratory of Particle Detection and Electronics, University of Science and Technology of China, Hefei, Anhui 230026, China} 
    \author{Ziyang Li}
    \affiliation{State Key Laboratory of Particle Detection and Electronics, University of Science and Technology of China, Hefei, Anhui 230026, China}
    \author{Rongrong Ma}
    \affiliation{Physics Department, Brookhaven National Laboratory, Upton, NY 11973, USA}
    \author{Zebo Tang}  
    \email{zbtang@ustc.edu.cn (Corresponding author)}
    \affiliation{State Key Laboratory of Particle Detection and Electronics, University of Science and Technology of China, Hefei, Anhui 230026, China}
    \author{Prithwish Tribedy}
    \affiliation{Physics Department, Brookhaven National Laboratory, Upton, NY 11973, USA}
    \author{Chun Yuen Tsang}
    \affiliation{Department of Physics, Kent State University, Kent, OH 44242, USA}
    \author{Zhangbu Xu}
    \affiliation{Physics Department, Brookhaven National Laboratory, Upton, NY 11973, USA}
    \author{Wangmei Zha}
    \affiliation{State Key Laboratory of Particle Detection and Electronics, University of Science and Technology of China, Hefei, Anhui 230026, China}

\date{\today}

\begin{abstract}

Baryon numbers are carried by valence quarks in the standard QCD picture of the baryon structure, while some theory proposed an alternative baryon number carrier, a non-perturbative Y-shaped configuration of the gluon field, called the baryon junction in the 1970s. However, neither of the theories has been verified experimentally. It was recently suggested to search for the baryon junction by investigating the correlation of net-charge and net-baryon yields at midrapidity in heavy-ion collisions. This paper presents studies of such correlations in collisions of various heavy ions from Oxygen to Uranium with the UrQMD Monte Carlo model. The UrQMD model implements valence quark transport as the primary means of charge and baryon stopping at midrapidity. Detailed study are also carried out for isobaric $_{40}^{96}\mathrm{Zr}$ + $_{40}^{96}\mathrm{Zr}$ and $_{44}^{96}\mathrm{Ru}$ + $_{44}^{96}\mathrm{Ru}$ collisions. We found a universal trend of the charge stopping with respect to the baryon stopping, and that the charge stopping is always more than the baryon stopping. This study provides a model baseline in valence quark transport for what is expected in net-charge and net-baryon yields at midrapidity of relativistic heavy-ion collisions. 
\end{abstract}


\maketitle


\section{Introduction}
Relativistic heavy-ion collisions provide an unique laboratory to create a deconfined, hot and dense medium, called the quark gluon plasma (QGP), and study its properties \cite{Adams:2005dq,Arsene:2004fa,Back:2004je,Adcox:2004mh}. In such collisions, incoming nuclei can be stopped at midrapidity with their kinetic energies deposited for the QGP formation. Since nuclei are composed of baryons, this phenomenon is usually referred to as the baryon stopping \cite{Gyulassy:1997mz,NA49:1998gaz,BRAHMS:2009wlg}. Understanding the baryon stopping mechanism is of prime importance for studying the QGP as it provides the initial conditions for the QGP creation. 

Baryon stopping can be investigated via the rapidity distribution of the net-baryon number, {\it i.e.}, the difference between numbers of baryons and anti-baryons, in heavy-ion collisions. This is because the baryon number is a strictly conserved quantity, and has to originate from the colliding nuclei. Since only protons and neutrons are long-lived baryons and it is much more difficult to detect neutrons than protons, baryon stopping is therefore usually probed experimentally via net-proton (proton minus anti-proton) distribution. Rapidity distributions of net-protons have been measured at colliders, such as Alternating Gradient Synchrotron (AGS), Super Proton Synchrotron (SPS), Relativistic Heavy Ion Collider (RHIC) and Large Hadron Collider (LHC), covering a large range of center-of-mass energy per nucleon-nucleon pair (\sqrtsnn)~\cite{E917:2000spt,E802:1999hit,E877:1999qdc,NA49:1998gaz,BRAHMS:2009wlg,STAR:2008med,ALICE:2013mez}. At the AGS with low collision energies of $3.63 \le\sqrtsnn\le4.85$ GeV, a large amount of net-protons are observed at midrapidity, which rapidly decrease towards forward rapidities. This is close to the extreme scenario of full stopping. On the other hand, at the LHC energy of \sqrtsnn\ = 2.76 TeV, there are nearly vanishing net-protons measured at midrapidity, close to the the extreme scenario of full transparency. Heavy-ion collisions at RHIC sit between these two extreme cases, and a finite number of net-protons are observed at midrapidity, which increase slowly towards forward rapidity (up to $y\approx 3$), indicating high but still partial transparency. 

While general features of the baryon stopping have been established in heavy-ion collisions, the underlying mechanism has not been fully understood. One key ingredient is the identify of the baryon number carrier in a nucleus. Traditionally, valence quarks are assumed to carry the baryon number in the QCD. Event generators of heavy-ion collisions, such as the Heavy Ion Jet Interaction Generator (HIJING)~\cite{Wang:1991hta}, the Ultra relativistic Quantum Molecular Dynamics (UrQMD)~\cite{Bass:1998ca,Petersen:2008dd}, A Multi-Phase Transport (AMPT)~\cite{Zhang:1999bd,Lin:2004en}, implement valence quark stopping by assuming that a nucleon is composed of a tightly bounded diquark and a quark. However, none of the models can reproduce the measured net-proton yields at midrapidity out of the box, due to the difficulty of moving valence quarks from incoming nuclei over a large rapidity gap (more than 5 units at \sqrtsnn\ = 200 GeV) to midrapidity. Instead, the valence quark stopping in those models needs to be enhanced parametrically in order to match experimental data.

\begin{figure}[btph]
\begin{center}
\includegraphics[width=0.6\columnwidth]{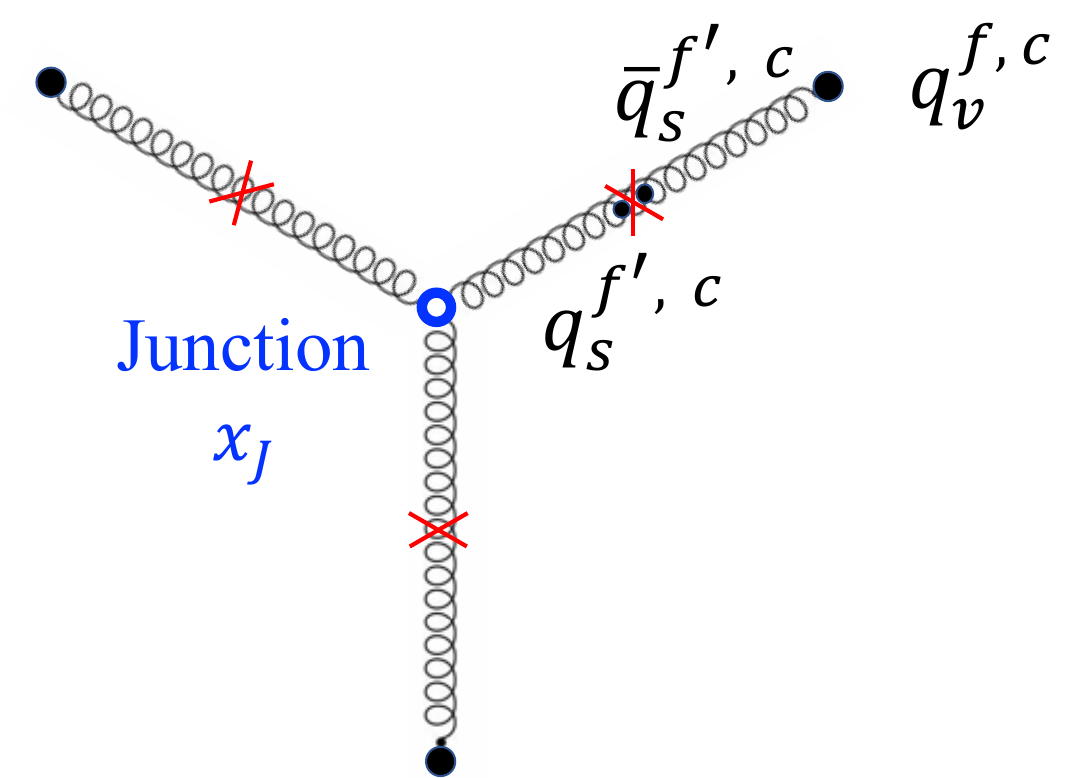}
\caption{Illustration of a baryon junction linking to three valence quarks in a baryon. The Wilson lines can be excited and fragment into $q\bar{q}$ sea quark pairs. The anti-sea-quark connected to a valence quark has the same color as the valence quark, but can have different flavor.}
\label{fig1}
\end{center}
\end{figure}

Alternatively, a ``baryon junction'' mechanism was proposed as the carrier of the baryon number~\cite{Artru:1974zn,Rossi:1977cy,Kharzeev:1996sq}. Some followup work can be found at~\cite{Vance:1997th, Vance:1999pr, ToporPop:2004lve, Bierlich:2022pfr, Shen:2022oyg, gyulassy2002proceedings,Kopeliovich:1998ps,arakelyan2002baryon,Takahashi:2000te}. Different from the traditional picture of the baryon structure, a baryon in this mechanism is composed of three valence quarks and a string junction linked to them by gluons, as illustrated schematically in Fig.~\ref{fig1}. 
The string junction is a non-perturbative Y-shaped configuration of gluon fields, which traces the baryon number instead of valence quarks. When the junction is pulled away from the baryon, the strings between the junction and valence quarks break up and $q\bar{q}$ pairs are produced. The resulting baryon is composed of three sea quarks with possibly different quark flavor content than the original baryon, while the valence quarks emerge as mesons. The baryon junction is expected to contain an infinite number of gluons, and therefore carries on average an infinitely small fraction of the baryon's momentum. Consequently, the baryon junction has a longer interaction time and thus more likely to be stopped at midrapidity compared to valence quarks, leading to enhanced baryon stopping.

Whether the stopped baryons come from valence quarks or baryon junctions could potentially be discriminated via the correlation of stopped baryons and charges at midrapidity in heavy-ion collisions~\cite{Brandenburg:2022hrp}, since the electric charge is carried by valence quarks. In the scenario of valence quark stopping, the stopped charges and baryons are expected to be proportional to the atomic number ($Z$) and mass number ($A$) of the colliding nuclei, respectively. In case of baryon junction stopping, stopped baryons are expected to scale with $A$, while the stopped charges are not correlated with $Z$. As illustrated in Fig.~\ref{fig1}, the stopped baryon is composed of three sea quarks from the fragmentation of the three Wilson lines linked to valence quarks. The three sea quarks can have different flavors from the valence quarks. Consequently, although the baryon number of the stopped baryon is the same as the the wounded baryon, their charges are not correlated.

In this article, correlations of baryon and charge stopping are studied using the UrQMD event generator, which employs valence quarks as the baryon number carriers. Such studies will serve as a baseline for the search of the baryon junction in heavy-ion experiments. The article is organized as the following. Section ~\ref{sec:method} introduces the UrQMD model, datasets and methods. Net-baryon and net-charge distributions and their correlations in collisions of various nucleus species will be discussed in Sec.~\ref{sec:results}.A and \ref{sec:results}.B, respectively. The differences between net-charge and net-baryon yields at midrapidity in isobaric Ru+Ru and Zr+Zr collisions will be shown in Sec.~\ref{sec:results}.C, and finally a summary is given in Sec.~\ref{sec:summary}.

\section{Model and Methodology}
\label{sec:method}
UrQMD is a microscopic transport model based on the covariant propagation of hadrons on classical trajectories in combination with stochastic binary scattering and resonance decays~\cite{Bass:1998ca,Petersen:2008dd}. It deals with both hadronic and partonic interactions through string formation and fragmentation. Cross sections of inelastic hadron-hadron interactions are either tabulated, or parameterized or extracted from other cross sections via general principles such as detailed balance or the additive quark model. As aforementioned, baryon stopping in UrQMD is realized by the excitation and fragmentation of the strings between diquark and quark. Incoherent multiple inelastic scatterings between the valence quark and diquark is the dominant mechanism to stop the baryons in colliding target and projectile, while the diquark breaking process is also included but of minor importance. It is worth noting that UrQMD uses a Gaussian parameterization for the longitudinal fragmentation, which was tuned to match net-proton and net-baryon rapidity distributions measured in heavy-ion collisions by enhancing valence quark stopping~\cite{Bleicher:1999xi,Bass:1998ca,Petersen:2008dd}.

In this study, collisions of $_8^{16}\mathrm{O}$  + $_8^{16}\mathrm{O}$, $_{13}^{27}\mathrm{Al}$ + $_{13}^{27}\mathrm{Al}$,  $_{29}^{64}\mathrm{Cul}$ + $_{29}^{64}\mathrm{Cu}$,  $_{40}^{96}\mathrm{Zr}$ + $_{40}^{96}\mathrm{Zr}$,  $_{44}^{96}\mathrm{Ru}$ + $_{44}^{96}\mathrm{Ru}$,  $_{79}^{197}\mathrm{Au}$ + $_{79}^{197}\mathrm{Au}$ and  $_{92}^{238}\mathrm{U}$ + $_{92}^{238}\mathrm{U}$ at \sqrtsnn\ = 200 GeV are simulated with UrQMD. The atomic number $Z$, the mass number $A$ and their ratios for the colliding nuclei are listed in Table~\ref{table}. 
\begin{table}[hptb]
\centering
 \caption{Atomic number ($Z$), mass number ($A$) and their ratios for various nuclei under study.}
   \label{table}
   \begin{tabular}{cccc}
\hline \hline
Nucleus & $Z$ & $A$ & $Z/A$ \\
\hline
O & 8 & 16 & 0.500 \\
Al & 13 & 27 &  0.481 \\
Cu & 29 & 64 & 0.453\\
Zr & 40 & 96 & 0.417 \\
Ru & 44 & 96 &0.458\\
Au & 79 & 197 &0.401\\
U & 92 & 238 & 0.386\\
\hline \hline
\end{tabular}
\end{table}
There is almost an order of magnitude coverage in $Z$, and $Z$/$A$ varies by about 23\% among different nuclei. To indicate the collision geometry, centrality, determined based on the impact parameter, is employed with a central (peripheral) collision corresponding to small (large) impact parameter or large (small) nuclear overlap. The centrality can be quantified using the number of participating nucleons (\npart) in each heavy-ion collision.

The charge and baryon stopping are quantified with the net-charge and net-baryon numbers at midrapidity, respectively. The net-charge number ($Q$) is defined as the sum of charges in the unit of electron charge ($e$) for all final-state particles within the desired acceptance. The net-baryon number ($B$) is defined as the sum of signed baryon number of all baryons ($p$, $n$, $\Lambda$, $\Sigma$, $\Xi$, $\Omega$ and their antiparticles). For simplicity, $Q$ and $B$ denote the average net-charge number and net-baryon number over all analyzed events for each collision system of a given centrality class. When integrated over the full phase space, $Q$ and $B$ are found to be exactly twice the $Z$ and $A$ of the colliding nuclei, respectively. This confirms the implementation of the conservation laws for charge and baryon quantum numbers in the UrQMD.

\section{Results}\label{sec:results}
\subsection{Net-charge and net-baryon rapidity distributions}

\begin{figure}[tbph]
\begin{center}

\includegraphics[width=0.7\columnwidth]{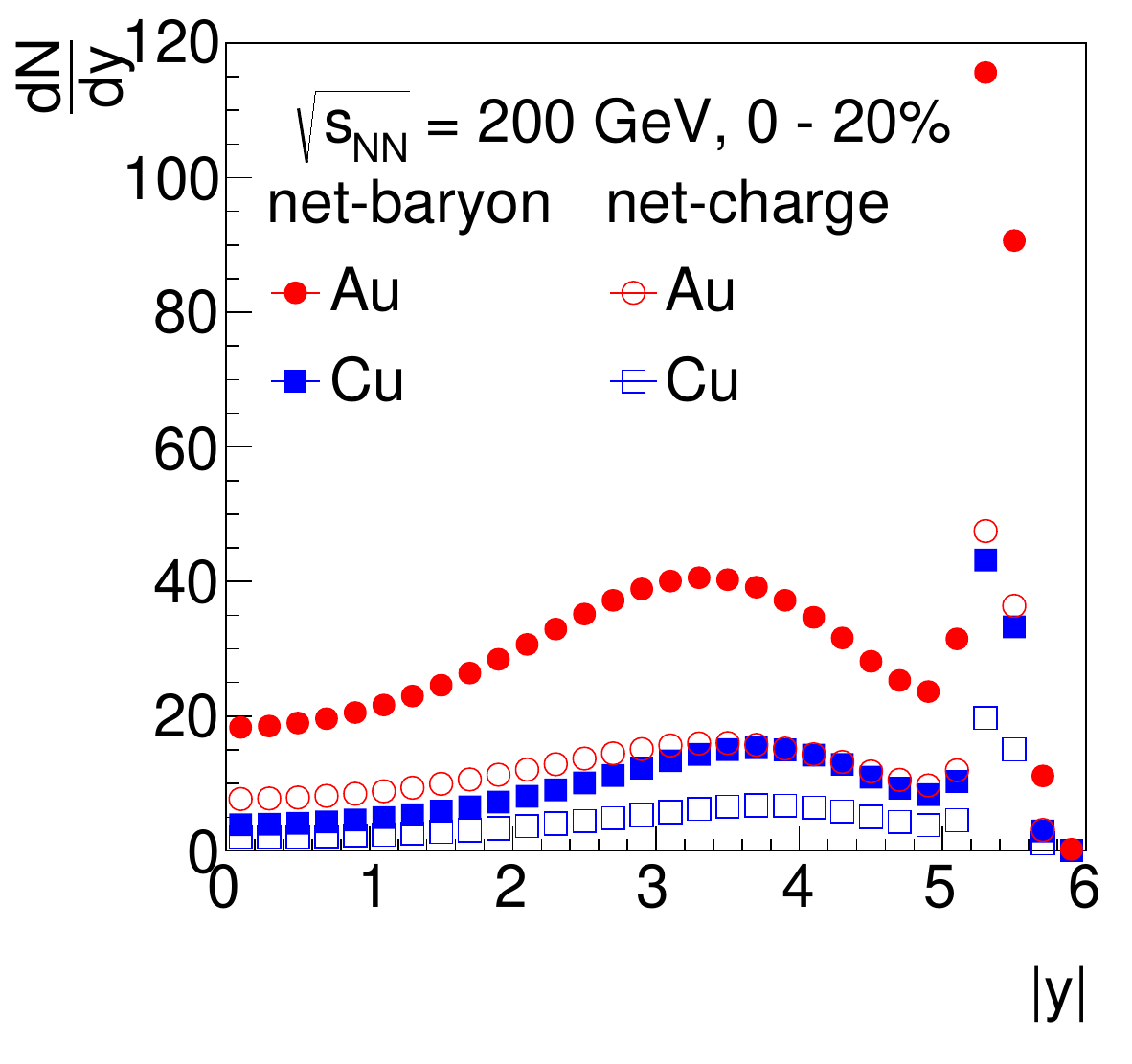}
\includegraphics[width=0.7\columnwidth]{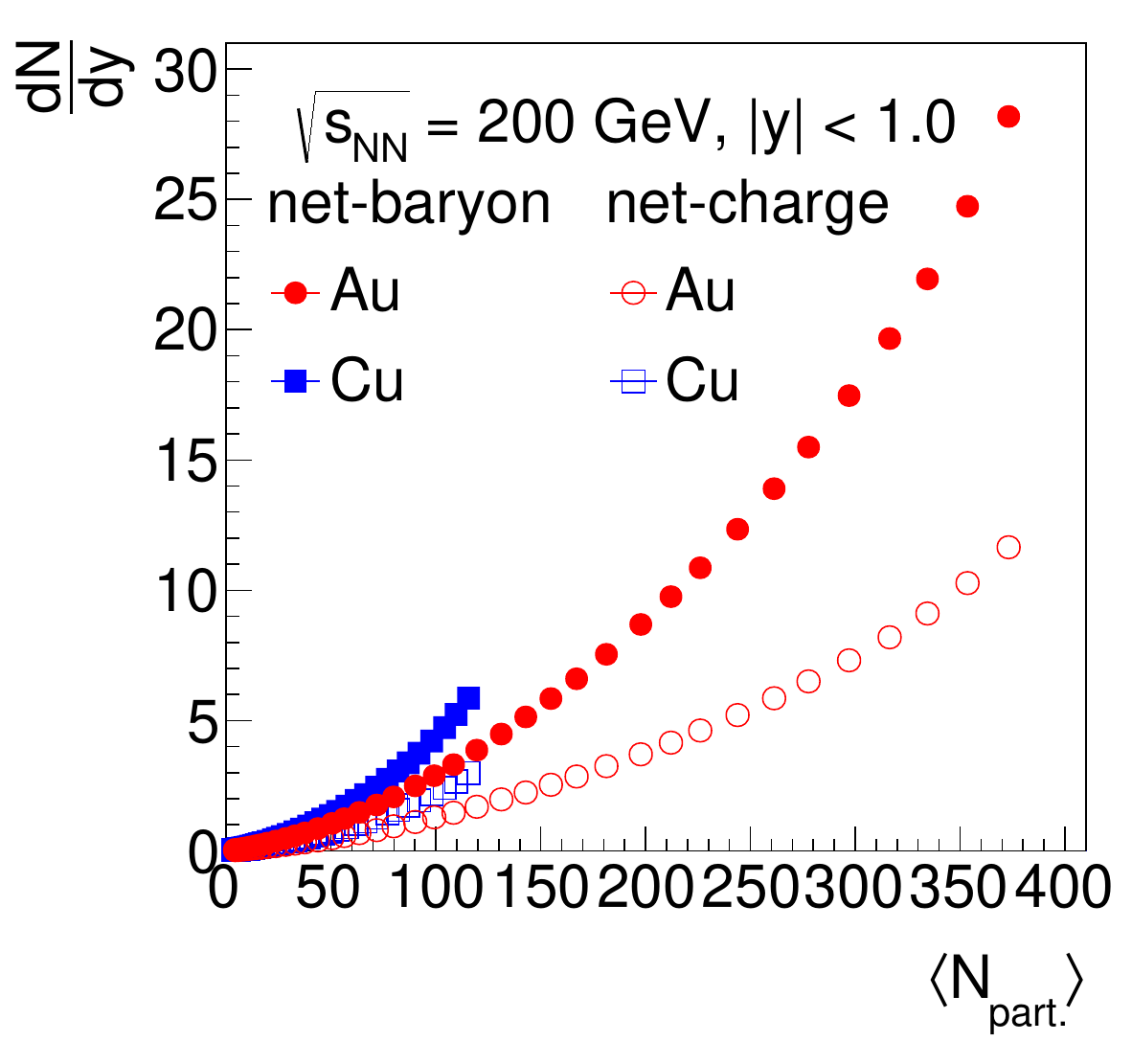}
\caption{\textbf{Top:} Net-charge and net-baryon rapidity distributions in 0-20\% central Au+Au and Cu+Cu collisions at \sqrtsnn\ = 200 GeV. \textbf{Bottom:} Net-charge and net-baryon rapidity density at midrapidity ($|y|<1.0$) as a function of \npart\ in Au+Au and Cu+Cu collisions. }
\label{fig2}
\end{center}
\end{figure}

Rapidity distributions of net-charge and net-baryon numbers in 0-20\% central Au+Au and Cu+Cu collisions at \sqrtsnn\ = 200 GeV are shown in the top panel of Fig.~\ref{fig2}.
Finite net-charges and net-baryons are stopped at midrapidity. The peak at $y\approx y_\mathrm{beam}=5.36$ corresponds to the nucleons not participating in the collisions (spectators). The rest of the net-charge and net-baryon distributions are peaked at $y \approx 3.4$ (3.8), about 2.0 (1.6) unit smaller than the beam rapidity, in Au+Au (Cu+Cu) collisions. The rapidity peak position for Au+Au collisions is about 0.4 unit smaller than that for Cu+Cu collisions, indicating system size dependence of charge and baryon stopping. The larger the collision size, the larger the stopping power. The bottom panel shows the net-charge and net-baryon densities at midrapidity ($|y|<1.0$) as a function of \npart. They increase dramatically from peripheral to central collisions. The net-charge and net-baryon yields in Cu+Cu collisions are higher than those in Au+Au collisions at the same \npart, likely due to different geometrical shapes of the overlapping region. At the same \npart, the mean thickness of the overlapping region in Cu+Cu collisions is larger than that in Au+Au collisions. Consequently, the partons in Cu+Cu collisions experience more scatterings on average, thus are more likely to be stopped at midrapidity.

\subsection{Correlations of net-charges and net-baryons}

\begin{figure}[tbph]
\begin{center}
\includegraphics[width=0.8\columnwidth]{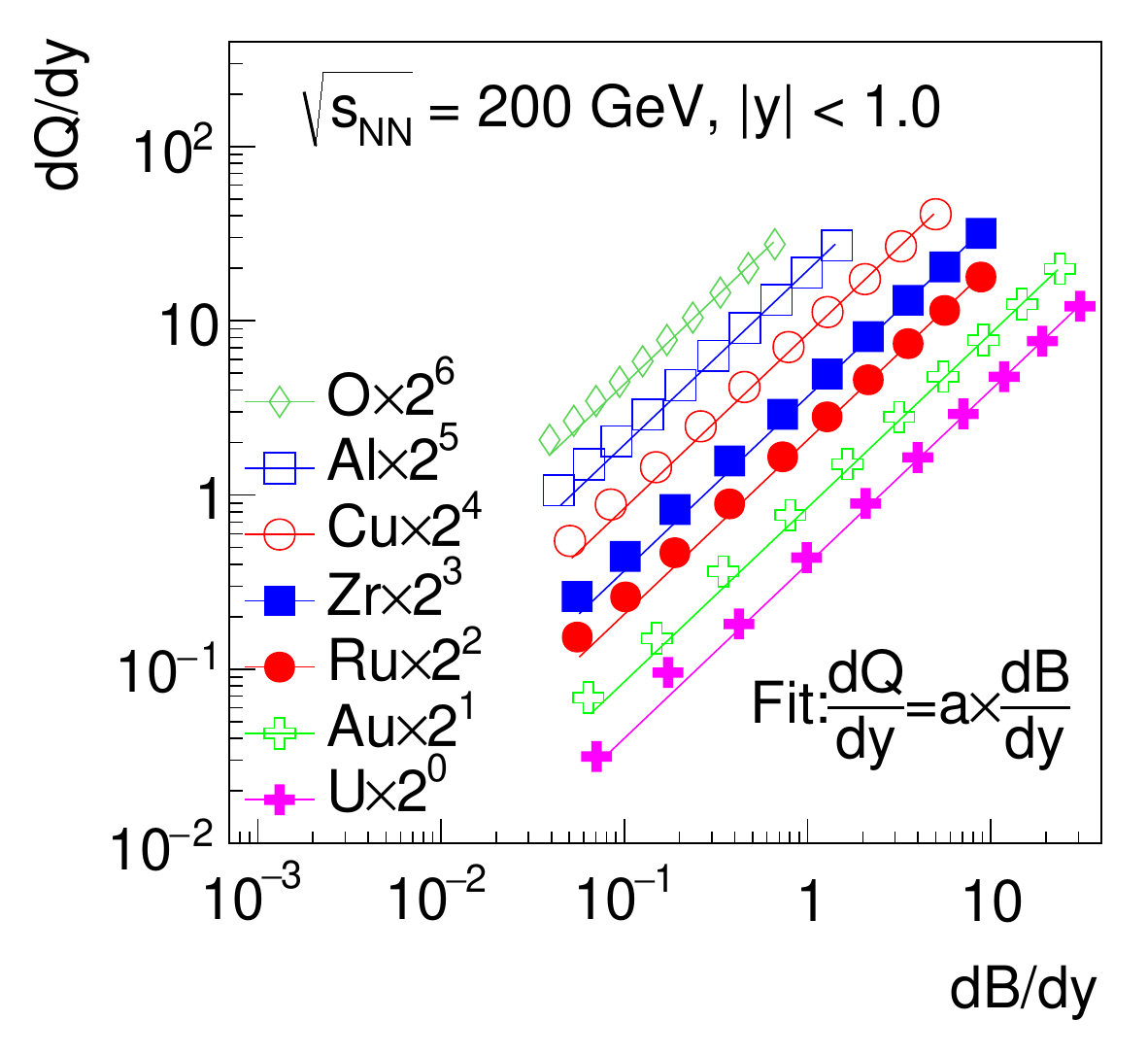}

\caption{Net-charge density as a function of net-baryon density at midrapidity in heavy-ion collisions of O+O, Al+Al, Cu+Cu, Zr+Zr, Ru+Ru, Au+Au and U+U at \sqrtsnn\ = 200 GeV. Data points are scaled up by factors of $2^n$ ($n = 1 - 6$) from Au to O for clarity. Solid curves indicate fit results to different sets of data points. }
\label{fig3}
\end{center}
\end{figure}

Figure~\ref{fig3} shows the correlations of net-charge ($dQ/dy$) and net-baryon ($dB/dy$) densities at midrapidity ($|y|<1.0$) in collisions of heavy ions with various $Z$ and $A$ at \sqrtsnn\ = 200 GeV. 
Centrality classes for each collision system are from 0 to 100\% with steps of 10\%. The leftmost data points correspond to the most peripheral collisions (90-100\%), while 0-10\% central collisions correspond to largest net-baryon densities. In all collision systems, the net-charge and net-baryon numbers are strongly correlated. To quantify the correlation, data points are fitted with
\begin{equation}
\frac{dQ}{dy} = a \times \frac{dB}{dy},
\end{equation}
where $a$ is the slope. The fitted results are shown as solid lines in Fig.~\ref{fig3}, which describe the data points in central collisions well but underestimate those in peripheral collisions. The different behaviors in central and peripheral collisions could be due to the effect of multiple scatterings, which grows stronger with larger collision system size. The slope $a$ increases from $0.401 \pm 0.001$ in U+U collisions to $0.677 \pm 0.001$ in O+O collisions, approximately proportional to $Z/A$. This is different from the expectation of baryon junction stopping, which predicts the slope to be around 0.5 and independent of $Z/A$, as the produced quarks linked to the stopped baryon junctions are from sea quarks.

\begin{figure}[tbph]
\begin{center}
\includegraphics[width=0.8\columnwidth]{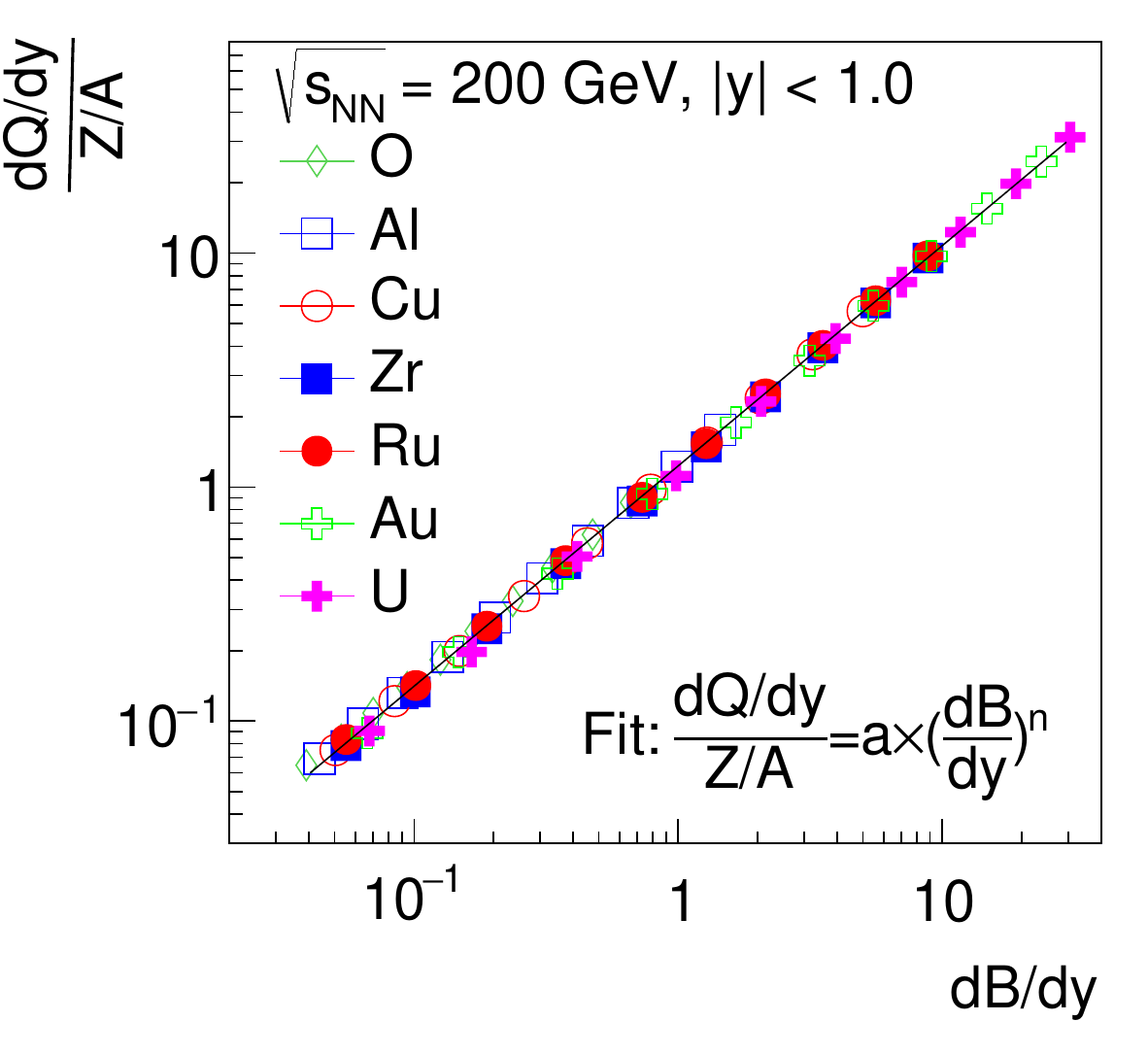}
\caption{$dQ/dy$, scaled by $A/Z$, versus $dB/dy$ at midrapidity ($|y|<1.0$) from different colliding systems. The solid line represents a simultaneous fit to all data points.}
\label{fig4}
\end{center}
\end{figure}

To further study the scaling behavior in different collision systems, distributions of $dQ/dy$, scaled by $A/Z$, are shown in Fig.~\ref{fig4} as a function of $dB/dy$. Results from O+O to U+U collisions collapse to a universal curve, which can be described by a power-law function: 
\begin{equation}
\frac{dQ/dy}{Z/A} = a \times (dB/dy)^n.
\end{equation}
The fitted parameters are: $a=1.2302 \pm 0.0007$ and $n=0.9413 \pm 0.0003$. 

\begin{figure}[tbp]
\begin{center}
\includegraphics[width=0.8\columnwidth]{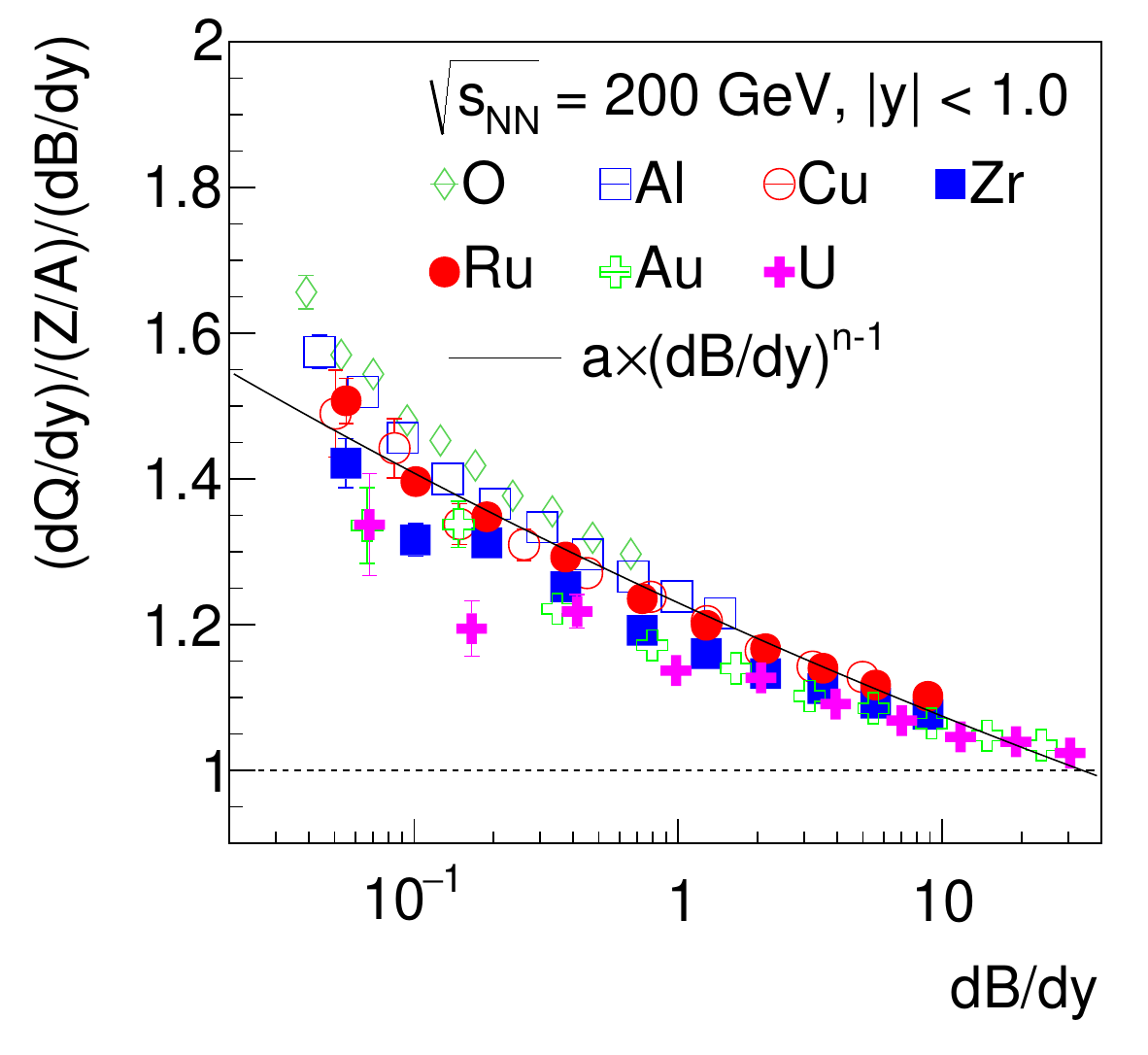}
\caption{The ratio of net-charge to net-baryon yield, scaled by $Z/A$, as a function of the net-baryon density within $|y|<1.0$. The solid line corresponds to the fitted curve shown in Fig.~\ref{fig4}. }
\label{fig:ratio}
\end{center}
\end{figure}

The ratios of $A/Z$-scaled $dQ/dy$ and $dB/dy$ as a function of $dB/dy$ in collisions of O+O to U+U are shown in Fig.~\ref{fig:ratio}. They follow a similar trend, being significantly larger than unity in small-system collisions with low net-baryon yields and approaching unity in large-system collisions with high net-baryon yields. A possible reason for the larger-than-unity ratio in small systems is that it is easier for transported valence quarks to form baryons at rapidities closer to the beam rapidity compared to at midrapidity. Thus mesons tend to have a flatter rapidity distribution than baryons, resulting in a larger net-charge to net-baryon yield ratio than $Z/A$ at midrapidity. This is more obvious for strange hadrons. In large systems, the multiple scattering effect is more pronounced and more quarks are present per unit of rapidity. These effects tend to wash out the difference in net-charge and net-baryon production against rapidity. The strong correlation of net-charge and net-baryon production from the valence quark transport provides a unique probe of baryon number carriers in nucleons with relativistic heavy ion collisions. The STAR Collaboration has collected large samples of Ru+Ru and Zr+Zr collisions in 2018, O+O collisions in 2021, Au+Au collisions in 2019 and 2023. These data could be used to test whether the ratio of net-charge and net-baryon at mid-rapidity is correlated with $Z/A$ or not.

\subsection{Net-charge and net-baryon differences between Ru+Ru and Zr+Zr collisions}

While examining the correlations between charge and baryon stopping at midrapidity provides a promising channel to discriminate between valence quarks and baryon junctions as the baryon number carriers, it is very difficult to measure the net-charge yield precisely in heavy-ion collisions. This is because realistic detectors have finite detection efficiencies and limited coverage in transverse momentum (\pT). For example, about 30\% of pions are missing below $\pT < 0.2$ GeV/$c$ in the measurement with the STAR experiment at RHIC \cite{STAR:2008med}. Corrections for these effects usually depend on particle species, and have non-negligible uncertainties. While such uncertainties on the particle yields are generally acceptable, they can render the net-charge measurement completely useless since the net-charge is calculated as the small difference between large yields of positive and negative particles. To overcome this difficulty, it was proposed to measure the net-charge difference between isobaric collisions of $_{40}^{96}\mathrm{Zr}$ + $_{40}^{96}\mathrm{Zr}$ and $_{44}^{96}\mathrm{Ru}$ + $_{44}^{96}\mathrm{Ru}$~\cite{Brandenburg:2022hrp}. This is motivated by that in 2018, the STAR experiment at RHIC recorded large samples of Ru+Ru and Zr+Zr collisions with almost identical running conditions~\cite{STAR:2021mii}. The net-charge difference ($\Delta Q = Q_{\rm{Ru}}-Q_{\rm{Zr}}$) can be calculated based on double ratios between positive and negative particles and between Ru+Ru and Zr+Zr collisions~\cite{Brandenburg:2022hrp}. Uncertainties in double ratios are negligible due to the cancellation of the uncertainties for different components of the double ratios. Consequently, one can compare $\Delta Q$ with $B\times \Delta Z/A$, where $\Delta Z = Z_{\rm{Ru}}-Z_{\rm{Zr}}$, and $B$ is the net-baryon number that is expected to be the same for the two isobaric collisions since the incoming nuclei carry the same baryon number. $\Delta Q$ should be close to $B\times \Delta Z/A$ in case of valence quark stopping, while for baryon junction stopping, $\Delta Q < B\times \Delta Z/A$ is expected.

\begin{figure}[tbp]
\begin{center}
\includegraphics[width=0.98\columnwidth]{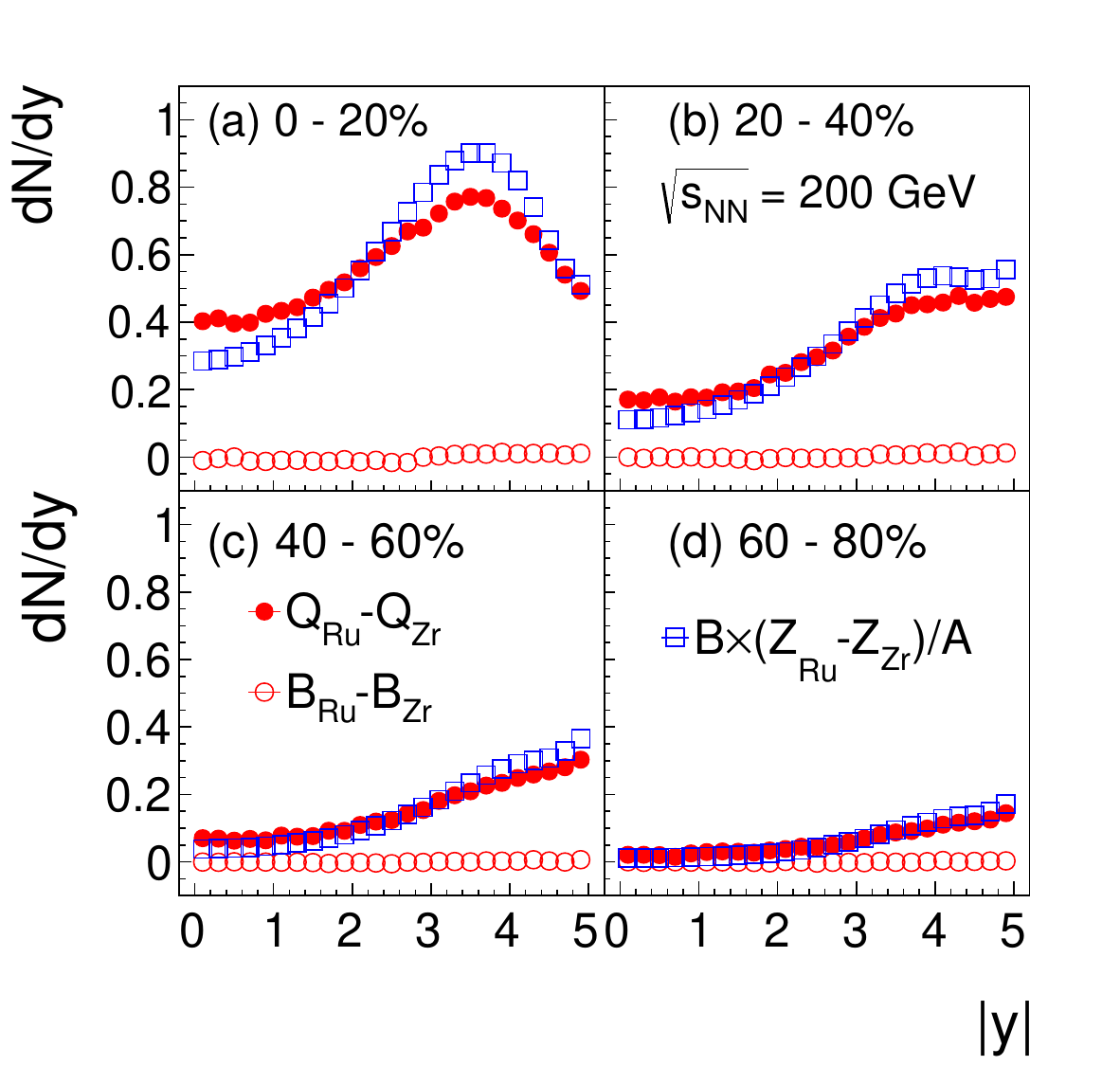}
\caption{Differences of $dQ/dy$ and $dB/dy$ between Ru+Ru and Zr+Zr collisions as a function of rapidity for 0-20\%, 20-40\%, 40-60\% and 60-80\% centralities. $dB/dy \times \Delta Z/A$ is also shown for comparison.}
\label{fig6}
\end{center}
\end{figure}

To provide the baseline for the baryon junction search using isobaric collisions, Fig.~\ref{fig6} shows $dQ/dy$ and $dB/dy$ differences as a function of rapidity between Ru+Ru and Zr+Zr collisions at \sqrtsnn\ = 200 GeV for 0-20\%, 20-40\%, 40-60\% and 60-80\% centralities. As expected, the net-baryon difference is consistent with 0 in Ru+Ru and Zr+Zr collisions (open circles). However, there are significantly more net-charges in Ru+Ru collisions than those in Zr+Zr collisions (filled circles) since the Ru nucleus carries 10\% more charges than the Zr nucleus. $dB/dy \times \Delta Z/A$ is also shown for comparison. The net-charge difference is close to $dB/dy \times \Delta Z/A$, but they exhibit slightly different rapidity dependence. The net-charge is larger than $dB/dy \times \Delta Z/A$ at midrapidity, and vice-versa at forward rapidity. The transition occurs at $y=2-3$.

\begin{figure}[tbp]
\begin{center}
\includegraphics[width=0.8\columnwidth]{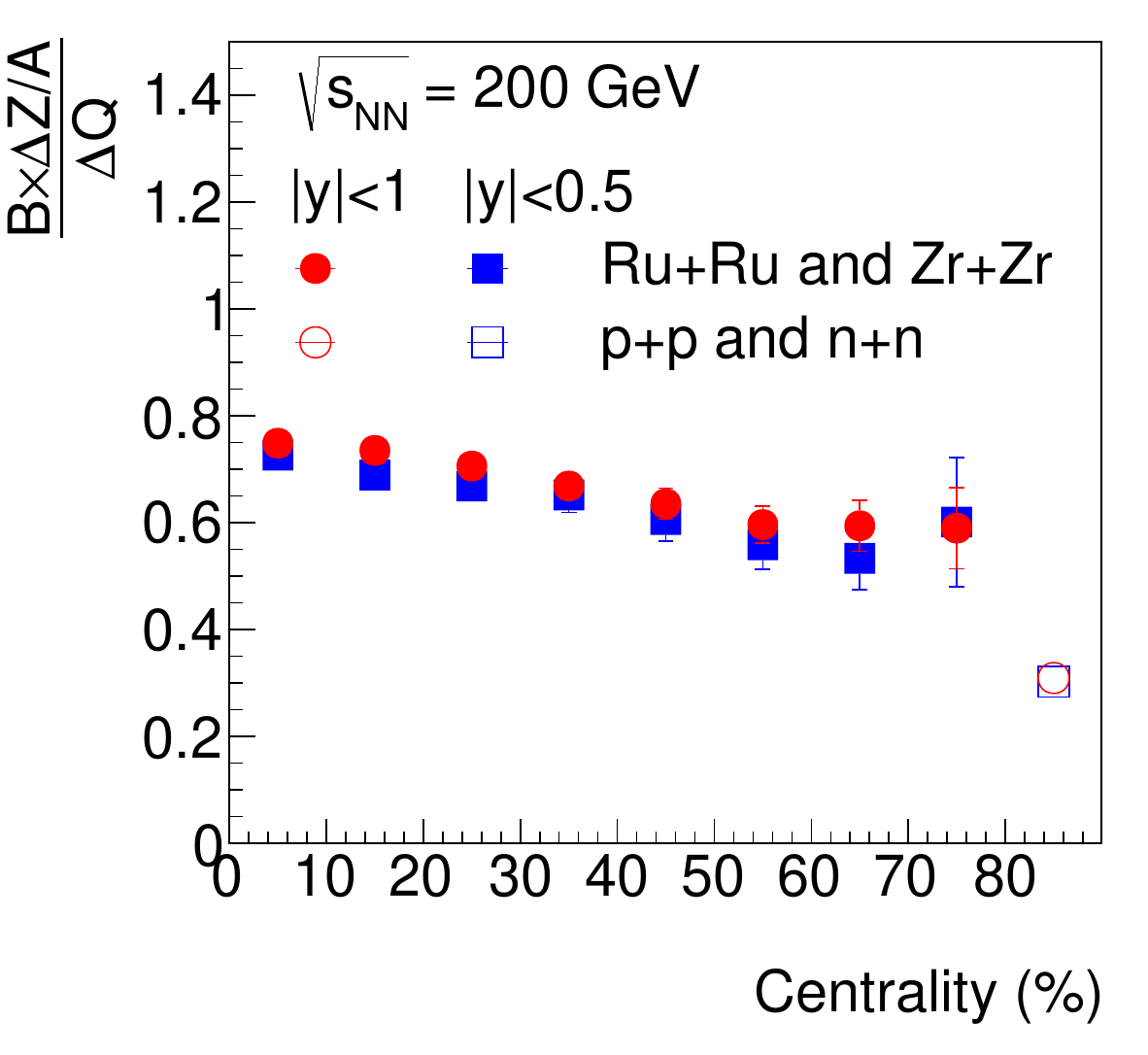}
\caption{($B\times \Delta Z/A)/ \Delta Q$ for $|y|<0.5$ and $|y|<1.0$ as a function of centrality in Ru+Ru and Zr+Zr collisions at \sqrtsnn\ = 200 GeV. Results from $p+p$ and $n+n$ collisions are also shown.}
\label{fig7}
\end{center}
\end{figure}

The ratios of $B \times \Delta Z/A$ over $\Delta Q$, integrated over midrapidity ($|y|<1.0$), are calculated in various centrality classes of 200 GeV Ru+Ru and Zr+Zr collisions and shown in Fig.~\ref{fig7}. They are below unity, mainly due to the different rapidity distributions of net-baryons and net-charges, as shown in Fig. \ref{fig6}. The effect is more obvious in peripheral collisions than in central collisions. Results for a narrower rapidity acceptance ($|y|<0.5$) are also shown. There are similar to those for $|y|<1.0$, indicating weak rapidity dependence within $|y|<1.0$ for such measurements. These results can be directly compared to experimental measurements in search for the baryon junction. The results from $p+p$ and $n+n$ collisions simulated with UrQMD at the same energy are also shown as open symbols, and they roughly follow the centrality dependence trend observed in Ru+Ru and Zr+Zr collisions. Recent STAR preliminary results show that the ratio of $B\times \Delta Z/A$  and $\Delta Q$ is significantly larger than unity in the isobaric collisions~\cite{LiYangThesis2023,TommyQM2023,NicoleDIS2023}.

\section{Summary}~\label{sec:summary}
In summary, the correlation between charge stopping and baryon stopping is studied in the collisions of various nuclei (from $_{8}^{16}\mathrm{O}$ to $_{92}^{238}\mathrm{U}$) at \sqrtsnn\ = 200 GeV with the UrQMD model, in which valence quarks carry the baryon number. It is found that the ratios of $A/Z$-scaled net-charge yield over net-baryon yield at midrapidity as a function of the net-baryon yield follow a universal trend in various collisions. In isobaric collisions ($_{44}^{96}\mathrm{Ru}$ + $_{44}^{96}\mathrm{Ru}$ and $_{40}^{96}\mathrm{Zr}$ + $_{40}^{96}\mathrm{Zr}$ ), the net-charges measured at midrapidity are correlated with the initial nuclear charge, and the ratios of $\Delta Z/A$-scaled net-baryon over the net-charge difference in the two isobaric collisions are finite and below unity. This provides an important baseline for the experimental search of the baryon junction in heavy-ion collisions. 

\section*{Ackonwledgement}
The authors would like to thank Drs. Declan Keane, Lijuan Ruan, James Daniel Brandenburg and Nicole Lewis for valuable discussions. This work is supported in part by the Strategic Priority Research Program of Chinese Academy of Sciences with Grant No. XDB34030000 and the National Key Research and Development Program of China under contract No. 2022YFA1604900, the U.S. DOE Office of Science365 under contract Nos. DE-SC0012704, DE-FG02-10ER41666, and DE-AC02-98CH10886.

\bibliography{ref}

\begin{thebibliography}{35}
\expandafter\ifx\csname natexlab\endcsname\relax\def\natexlab#1{#1}\fi
\providecommand{\url}[1]{\texttt{#1}}
\providecommand{\href}[2]{#2}
\providecommand{\path}[1]{#1}
\providecommand{\DOIprefix}{doi:}
\providecommand{\ArXivprefix}{arXiv:}
\providecommand{\URLprefix}{URL: }
\providecommand{\Pubmedprefix}{pmid:}
\providecommand{\doi}[1]{\href{http://dx.doi.org/#1}{\path{#1}}}
\providecommand{\Pubmed}[1]{\href{pmid:#1}{\path{#1}}}
\providecommand{\bibinfo}[2]{#2}
\ifx\xfnm\relax \def\xfnm[#1]{\unskip,\space#1}\fi
\bibitem[{Adams et~al.(2005)}]{Adams:2005dq}
\bibinfo{author}{J.~Adams}, et~al. (\bibinfo{collaboration}{STAR}),
\newblock \bibinfo{title}{{Experimental and theoretical challenges in the
  search for the quark gluon plasma: The STAR Collaboration's critical
  assessment of the evidence from RHIC collisions}},
\newblock \bibinfo{journal}{Nucl. Phys.} \bibinfo{volume}{A757}
  (\bibinfo{year}{2005}) \bibinfo{pages}{102--183}.
\bibitem[{Arsene et~al.(2005)}]{Arsene:2004fa}
\bibinfo{author}{I.~Arsene}, et~al. (\bibinfo{collaboration}{BRAHMS}),
\newblock \bibinfo{title}{{Quark gluon plasma and color glass condensate at
  RHIC? The Perspective from the BRAHMS experiment}},
\newblock \bibinfo{journal}{Nucl. Phys.} \bibinfo{volume}{A757}
  (\bibinfo{year}{2005}) \bibinfo{pages}{1--27}.
\bibitem[{Back et~al.(2005)}]{Back:2004je}
\bibinfo{author}{B.~B. Back}, et~al.,
\newblock \bibinfo{title}{{The PHOBOS perspective on discoveries at RHIC}},
\newblock \bibinfo{journal}{Nucl. Phys.} \bibinfo{volume}{A757}
  (\bibinfo{year}{2005}) \bibinfo{pages}{28--101}.
\bibitem[{Adcox et~al.(2005)}]{Adcox:2004mh}
\bibinfo{author}{K.~Adcox}, et~al. (\bibinfo{collaboration}{PHENIX}),
\newblock \bibinfo{title}{{Formation of dense partonic matter in relativistic
  nucleus-nucleus collisions at RHIC: Experimental evaluation by the PHENIX
  collaboration}},
\newblock \bibinfo{journal}{Nucl. Phys.} \bibinfo{volume}{A757}
  (\bibinfo{year}{2005}) \bibinfo{pages}{184--283}.
\bibitem[{Gyulassy et~al.(1997)Gyulassy, Topor~Pop, and
  Vance}]{Gyulassy:1997mz}
\bibinfo{author}{M.~Gyulassy}, \bibinfo{author}{V.~Topor~Pop},
  \bibinfo{author}{S.~E. Vance},
\newblock \bibinfo{title}{{Baryon number transport in high-energy nuclear
  collisions}},
\newblock \bibinfo{journal}{Acta Phys. Hung. A} \bibinfo{volume}{5}
  (\bibinfo{year}{1997}) \bibinfo{pages}{299--318}.
\bibitem[{Appelshauser et~al.(1999)}]{NA49:1998gaz}
\bibinfo{author}{H.~Appelshauser}, et~al. (\bibinfo{collaboration}{NA49}),
\newblock \bibinfo{title}{{Baryon stopping and charged particle distributions
  in central Pb + Pb collisions at 158-GeV per nucleon}},
\newblock \bibinfo{journal}{Phys. Rev. Lett.} \bibinfo{volume}{82}
  (\bibinfo{year}{1999}) \bibinfo{pages}{2471--2475}.
\bibitem[{Arsene et~al.(2009)}]{BRAHMS:2009wlg}
\bibinfo{author}{I.~C. Arsene}, et~al. (\bibinfo{collaboration}{BRAHMS}),
\newblock \bibinfo{title}{{Nuclear stopping and rapidity loss in Au+Au
  collisions at s(NN)**(1/2) = 62.4-GeV}},
\newblock \bibinfo{journal}{Phys. Lett. B} \bibinfo{volume}{677}
  (\bibinfo{year}{2009}) \bibinfo{pages}{267--271}.
\bibitem[{Back et~al.(2001)}]{E917:2000spt}
\bibinfo{author}{B.~B. Back}, et~al. (\bibinfo{collaboration}{E917}),
\newblock \bibinfo{title}{{Baryon rapidity loss in relativistic Au+Au
  collisions}},
\newblock \bibinfo{journal}{Phys. Rev. Lett.} \bibinfo{volume}{86}
  (\bibinfo{year}{2001}) \bibinfo{pages}{1970--1973}.
\bibitem[{Ahle et~al.(1999)}]{E802:1999hit}
\bibinfo{author}{L.~Ahle}, et~al. (\bibinfo{collaboration}{E802}),
\newblock \bibinfo{title}{{Proton and deuteron production in Au + Au reactions
  at 11.6/A-GeV/c}},
\newblock \bibinfo{journal}{Phys. Rev. C} \bibinfo{volume}{60}
  (\bibinfo{year}{1999}) \bibinfo{pages}{064901}.
\bibitem[{Barrette et~al.(2000)}]{E877:1999qdc}
\bibinfo{author}{J.~Barrette}, et~al. (\bibinfo{collaboration}{E877}),
\newblock \bibinfo{title}{{Proton and pion production in Au + Au collisions at
  10.8A-GeV/c}},
\newblock \bibinfo{journal}{Phys. Rev. C} \bibinfo{volume}{62}
  (\bibinfo{year}{2000}) \bibinfo{pages}{024901}.
\bibitem[{Abelev et~al.(2009)}]{STAR:2008med}
\bibinfo{author}{B.~I. Abelev}, et~al. (\bibinfo{collaboration}{STAR}),
\newblock \bibinfo{title}{{Systematic Measurements of Identified Particle
  Spectra in $p p, d^+$ Au and Au+Au Collisions from STAR}},
\newblock \bibinfo{journal}{Phys. Rev. C} \bibinfo{volume}{79}
  (\bibinfo{year}{2009}) \bibinfo{pages}{034909}.
\bibitem[{Abelev et~al.(2013)}]{ALICE:2013mez}
\bibinfo{author}{B.~Abelev}, et~al. (\bibinfo{collaboration}{ALICE}),
\newblock \bibinfo{title}{{Centrality dependence of $\pi$, K, p production in
  Pb-Pb collisions at $\sqrt{s_{NN}}$ = 2.76 TeV}},
\newblock \bibinfo{journal}{Phys. Rev. C} \bibinfo{volume}{88}
  (\bibinfo{year}{2013}) \bibinfo{pages}{044910}.
\bibitem[{Wang and Gyulassy(1991)}]{Wang:1991hta}
\bibinfo{author}{X.-N. Wang}, \bibinfo{author}{M.~Gyulassy},
\newblock \bibinfo{title}{{HIJING: A Monte Carlo model for multiple jet
  production in p p, p A and A A collisions}},
\newblock \bibinfo{journal}{Phys. Rev. D} \bibinfo{volume}{44}
  (\bibinfo{year}{1991}) \bibinfo{pages}{3501--3516}.
\bibitem[{Bass et~al.(1998)}]{Bass:1998ca}
\bibinfo{author}{S.~A. Bass}, et~al.,
\newblock \bibinfo{title}{{Microscopic models for ultrarelativistic heavy ion
  collisions}},
\newblock \bibinfo{journal}{Prog. Part. Nucl. Phys.} \bibinfo{volume}{41}
  (\bibinfo{year}{1998}) \bibinfo{pages}{255--369}.
\bibitem[{Petersen et~al.(2008)Petersen, Steinheimer, Burau, Bleicher, and
  St\"ocker}]{Petersen:2008dd}
\bibinfo{author}{H.~Petersen}, \bibinfo{author}{J.~Steinheimer},
  \bibinfo{author}{G.~Burau}, \bibinfo{author}{M.~Bleicher},
  \bibinfo{author}{H.~St\"ocker},
\newblock \bibinfo{title}{{A Fully Integrated Transport Approach to Heavy Ion
  Reactions with an Intermediate Hydrodynamic Stage}},
\newblock \bibinfo{journal}{Phys. Rev. C} \bibinfo{volume}{78}
  (\bibinfo{year}{2008}) \bibinfo{pages}{044901}.
\bibitem[{Zhang et~al.(2000)Zhang, Ko, Li, and Lin}]{Zhang:1999bd}
\bibinfo{author}{B.~Zhang}, \bibinfo{author}{C.~M. Ko}, \bibinfo{author}{B.-A.
  Li}, \bibinfo{author}{Z.-w. Lin},
\newblock \bibinfo{title}{{A multiphase transport model for nuclear collisions
  at RHIC}},
\newblock \bibinfo{journal}{Phys. Rev. C} \bibinfo{volume}{61}
  (\bibinfo{year}{2000}) \bibinfo{pages}{067901}.
\bibitem[{Lin et~al.(2005)Lin, Ko, Li, Zhang, and Pal}]{Lin:2004en}
\bibinfo{author}{Z.-W. Lin}, \bibinfo{author}{C.~M. Ko}, \bibinfo{author}{B.-A.
  Li}, \bibinfo{author}{B.~Zhang}, \bibinfo{author}{S.~Pal},
\newblock \bibinfo{title}{{A Multi-phase transport model for relativistic heavy
  ion collisions}},
\newblock \bibinfo{journal}{Phys. Rev. C} \bibinfo{volume}{72}
  (\bibinfo{year}{2005}) \bibinfo{pages}{064901}.
\bibitem[{Artru(1975)}]{Artru:1974zn}
\bibinfo{author}{X.~Artru},
\newblock \bibinfo{title}{{String Model with Baryons: Topology, Classical
  Motion}},
\newblock \bibinfo{journal}{Nucl. Phys. B} \bibinfo{volume}{85}
  (\bibinfo{year}{1975}) \bibinfo{pages}{442--460}.
\bibitem[{Rossi and Veneziano(1977)}]{Rossi:1977cy}
\bibinfo{author}{G.~C. Rossi}, \bibinfo{author}{G.~Veneziano},
\newblock \bibinfo{title}{{A Possible Description of Baryon Dynamics in Dual
  and Gauge Theories}},
\newblock \bibinfo{journal}{Nucl. Phys. B} \bibinfo{volume}{123}
  (\bibinfo{year}{1977}) \bibinfo{pages}{507--545}.
\bibitem[{Kharzeev(1996)}]{Kharzeev:1996sq}
\bibinfo{author}{D.~Kharzeev},
\newblock \bibinfo{title}{{Can gluons trace baryon number?}},
\newblock \bibinfo{journal}{Phys. Lett. B} \bibinfo{volume}{378}
  (\bibinfo{year}{1996}) \bibinfo{pages}{238--246}.
\bibitem[{Vance et~al.(1998)Vance, Gyulassy, and Wang}]{Vance:1997th}
\bibinfo{author}{S.~E. Vance}, \bibinfo{author}{M.~Gyulassy},
  \bibinfo{author}{X.~N. Wang},
\newblock \bibinfo{title}{{Baryon junction stopping at the SPS and RHIC via
  HIJING/B}},
\newblock \bibinfo{journal}{Nucl. Phys. A} \bibinfo{volume}{638}
  (\bibinfo{year}{1998}) \bibinfo{pages}{395C--398C}.
\bibitem[{Vance and Gyulassy(1999)}]{Vance:1999pr}
\bibinfo{author}{S.~E. Vance}, \bibinfo{author}{M.~Gyulassy},
\newblock \bibinfo{title}{{Anti-hyperon enhancement through baryon junction
  loops}},
\newblock \bibinfo{journal}{Phys. Rev. Lett.} \bibinfo{volume}{83}
  (\bibinfo{year}{1999}) \bibinfo{pages}{1735--1738}.
\bibitem[{Topor~Pop et~al.(2004)Topor~Pop, Gyulassy, Barrette, Gale, Wang, and
  Xu}]{ToporPop:2004lve}
\bibinfo{author}{V.~Topor~Pop}, \bibinfo{author}{M.~Gyulassy},
  \bibinfo{author}{J.~Barrette}, \bibinfo{author}{C.~Gale},
  \bibinfo{author}{X.~N. Wang}, \bibinfo{author}{N.~Xu},
\newblock \bibinfo{title}{{Baryon junction loops in HIJING / B anti-B v2.0 and
  the baryon /meson anomaly at RHIC}},
\newblock \bibinfo{journal}{Phys. Rev. C} \bibinfo{volume}{70}
  (\bibinfo{year}{2004}) \bibinfo{pages}{064906}.
\bibitem[{Bierlich et~al.(2022)}]{Bierlich:2022pfr}
\bibinfo{author}{C.~Bierlich}, et~al.,
\newblock \bibinfo{title}{{A comprehensive guide to the physics and usage of
  PYTHIA 8.3}}  (\bibinfo{year}{2022}).
\bibitem[{Shen and Schenke(2022)}]{Shen:2022oyg}
\bibinfo{author}{C.~Shen}, \bibinfo{author}{B.~Schenke},
\newblock \bibinfo{title}{{Longitudinal dynamics and particle production in
  relativistic nuclear collisions}},
\newblock \bibinfo{journal}{Phys. Rev. C} \bibinfo{volume}{105}
  (\bibinfo{year}{2022}) \bibinfo{pages}{064905}.
\bibitem[{Organizers: et~al.(2002)Organizers:, Gyulassy, Kharzeev, and
  Xu}]{gyulassy2002proceedings}
\bibinfo{author}{Organizers:}, \bibinfo{author}{M.~Gyulassy},
  \bibinfo{author}{D.~Kharzeev}, \bibinfo{author}{N.~Xu},
  \bibinfo{title}{{PROCEEDINGS OF RIKEN BNL RESEARCH CENTER WORKSHOP ON BARYON
  DYNAMICS AT RHIC, \url{https://www.bnl.gov/isd/documents/24266.pdf}}},
  \bibinfo{year}{2002}.
\bibitem[{Kopeliovich and Povh(1999)}]{Kopeliovich:1998ps}
\bibinfo{author}{B.~Kopeliovich}, \bibinfo{author}{B.~Povh},
\newblock \bibinfo{title}{{Baryon stopping at HERA: Evidence for gluonic
  mechanism}},
\newblock \bibinfo{journal}{Phys. Lett. B} \bibinfo{volume}{446}
  (\bibinfo{year}{1999}) \bibinfo{pages}{321--325}.
\bibitem[{Arakelyan et~al.(2002)Arakelyan, Capella, Kaidalov, and
  Shabelski}]{arakelyan2002baryon}
\bibinfo{author}{G.~Arakelyan}, \bibinfo{author}{A.~Capella},
  \bibinfo{author}{A.~Kaidalov}, \bibinfo{author}{Y.~M. Shabelski},
\newblock \bibinfo{title}{Baryon number transfer in hadronic interactions},
\newblock \bibinfo{journal}{The European Physical Journal C-Particles and
  Fields} \bibinfo{volume}{26} (\bibinfo{year}{2002}) \bibinfo{pages}{81--90}.
\bibitem[{Takahashi et~al.(2001)Takahashi, Matsufuru, Nemoto, and
  Suganuma}]{Takahashi:2000te}
\bibinfo{author}{T.~T. Takahashi}, \bibinfo{author}{H.~Matsufuru},
  \bibinfo{author}{Y.~Nemoto}, \bibinfo{author}{H.~Suganuma},
\newblock \bibinfo{title}{{The Three quark potential in the SU(3) lattice
  QCD}},
\newblock \bibinfo{journal}{Phys. Rev. Lett.} \bibinfo{volume}{86}
  (\bibinfo{year}{2001}) \bibinfo{pages}{18--21}.
\bibitem[{Brandenburg et~al.(2022)Brandenburg, Lewis, Tribedy, and
  Xu}]{Brandenburg:2022hrp}
\bibinfo{author}{J.~D. Brandenburg}, \bibinfo{author}{N.~Lewis},
  \bibinfo{author}{P.~Tribedy}, \bibinfo{author}{Z.~Xu},
\newblock \bibinfo{title}{{Search for baryon junctions in photonuclear
  processes and isobar collisions at RHIC}}  (\bibinfo{year}{2022}).
\bibitem[{Bleicher et~al.(1999)}]{Bleicher:1999xi}
\bibinfo{author}{M.~Bleicher}, et~al.,
\newblock \bibinfo{title}{{Relativistic hadron hadron collisions in the
  ultrarelativistic quantum molecular dynamics model}},
\newblock \bibinfo{journal}{J. Phys. G} \bibinfo{volume}{25}
  (\bibinfo{year}{1999}) \bibinfo{pages}{1859--1896}.
\bibitem[{Abdallah et~al.(2022)}]{STAR:2021mii}
\bibinfo{author}{M.~Abdallah}, et~al. (\bibinfo{collaboration}{STAR}),
\newblock \bibinfo{title}{{Search for the chiral magnetic effect with isobar
  collisions at $\sqrt {s_{NN}}$=200 GeV by the STAR Collaboration at the BNL
  Relativistic Heavy Ion Collider}},
\newblock \bibinfo{journal}{Phys. Rev. C} \bibinfo{volume}{105}
  (\bibinfo{year}{2022}) \bibinfo{pages}{014901}.
\bibitem[{Li(2023)}]{LiYangThesis2023}
\bibinfo{author}{Y.~Li}, \bibinfo{title}{Measurement of identified particle
  spectra and tracking the baryon number in Ru+Ru and Zr+Zr collisions at
  $\sqrt{s_{NN}}$ = 200 GeV}, Ph.D. thesis, University of Science and
  Technology of China, \bibinfo{year}{2023}.
\bibitem[{Tsang(2023)}]{TommyQM2023}
\bibinfo{author}{C.~Y. Tsang} (\bibinfo{collaboration}{for the STAR
  Collaboration}),
\newblock \bibinfo{title}{Search for evidence of the baryon junction in
  photonuclear processes and heavy-ion collisions at star},
\newblock in: \bibinfo{booktitle}{XXXth International Conference on
  Ultra-relativistic Nucleus-Nucleus Collisions (Quark Matter 2023)},
  \bibinfo{year}{2023}.
\bibitem[{Lewis(2023)}]{NicoleDIS2023}
\bibinfo{author}{N.~Lewis} (\bibinfo{collaboration}{for the STAR
  Collaboration}),
\newblock \bibinfo{title}{Search for baryon junctions in photonuclear processes
  and isobar collisions at star},
\newblock in: \bibinfo{booktitle}{The 30th International Workshop on Deep
  Inelastic Scattering}, \bibinfo{year}{2023}.

\end{thebibliography}

\end{document}